\documentclass[12pt]{article}
\usepackage{geometry}                
\usepackage{graphicx}
\usepackage{amssymb}
\usepackage{amsmath}
\usepackage{epstopdf}
\usepackage{verbatim}

\def\ba{\begin{array}}
\def\ea{\end{array}}

\def\td{\tilde}

\def\dalemb#1#2{{\vbox{\hrule height .#2pt
       \hbox{\vrule width.#2pt height#1pt \kern#1pt
               \vrule width.#2pt}
       \hrule height.#2pt}}}

\newcommand{\Tr}{{\rm Tr} }
\def\ep{{\epsilon}}

\def\rh{{\tilde{r}}}
\def\th{{\tilde{t}}}
\def\phih{{\tilde{\phi}}}

\newcommand{\be}{\begin{equation}}
\newcommand{\ee}{\end{equation}}
\newcommand{\bal}{\begin{align}}
 \newcommand{\eal}{\end{align}}
\newcommand{\ben}{\begin{equation*}}
\newcommand{\een}{\end{equation*}}
\newcommand{\bea}{\begin{eqnarray}}
\newcommand{\eea}{\end{eqnarray}}
\newcommand{\bean}{\begin{eqnarray*}}
\newcommand{\eean}{\end{eqnarray*}}
\newcommand{\bes}{\begin{subequations}}
\newcommand{\ees}{\end{subequations}}

\begin{document}

\begin{titlepage}
\bigskip
\rightline{}

\bigskip\bigskip\bigskip\bigskip
\centerline {\Large \bf {Uniqueness of Extremal Kerr and  }}
\bigskip
\centerline{\Large\bf {Kerr-Newman Black Holes}}
\bigskip\bigskip
\bigskip\bigskip

\centerline{\large  Aaron J. Amsel, Gary T. Horowitz, Donald Marolf, and Matthew M. Roberts}
\bigskip\bigskip
\centerline{\em Department of Physics, UCSB, Santa Barbara, CA 93106}
\centerline{\em  amsel@physics.ucsb.edu, gary@physics.ucsb.edu}
\centerline{\em marolf@physics.ucsb.edu, matt@physics.ucsb.edu}
\bigskip\bigskip

\begin{abstract}
We prove that the only four dimensional, stationary, rotating, asymptotically flat (analytic) vacuum  black hole with a single degenerate horizon is given by the extremal Kerr  solution. We also prove a similar uniqueness theorem for the extremal Kerr-Newman solution. This closes a longstanding gap in the black hole uniqueness theorems.
\end{abstract}
\end{titlepage}

\bigskip
\baselineskip 16pt

\setcounter{equation}{0}

\section{Introduction}

In the 1970's, work by Hawking \cite{Hawking}, Carter \cite{Carter}, and Robinson \cite{Robinson:1975bv} proved that the  only stationary, asymptotically flat vacuum black hole with a (single) non-degenerate horizon is the nonextremal Kerr metric. In the early 1980's, Mazur \cite{Mazur:1982db} and Bunting \cite{Bunting} extended this proof to the charged Kerr-Newman black hole. These uniqueness theorems have been the basis for most of the subsequent work on black holes for almost thirty years. This spans a wide range of topics from astrophysical black holes to no hair theorems to studies of black hole thermodynamics and quantum aspects of black holes. 

A similar uniqueness theorem for the extremal Kerr or Kerr-Newman black hole has not been available. The existing techniques were not sufficient to obtain a proof in this case.
Two recent developments have encouraged us to reexamine this longstanding problem.
First, it was shown that stationary rotating (analytic) extremal black holes must be axisymmetric \cite{Hollands:2008wn,Chrusciel:2008js}. This extended the well known result for nonextremal black holes \cite{Hawking,Sudarsky:1992ty,Chrusciel:1993cv,Friedrich:1998wq} to the extremal case. Second,
  building on earlier work \cite{Kunduri:2007vf}, it has recently been shown that the near horizon geometry of any extremal vacuum black hole must agree with the extremal  Kerr metric \cite{Kunduri:2008tk}. Similarly, the near horizon geometry of any extremal electrovac black hole must agree with the extremal Kerr-Newman solution \cite{Kunduri:2008tk}.\footnote{See \cite{Hajicek,Lewandowski:2002ua} for  related results.} At first sight, these local uniqueness theorems  seem surprising since one might expect that adding stationary matter outside the black hole could distort the horizon. However, extremal horizons are infinitely far away from any matter outside and do not get distorted. We will show that these new results can be combined  with existing methods of proving black hole uniqueness to finally prove the uniqueness of the extremal Kerr and Kerr-Newman solution.

\section{Uniqueness of the Extremal Kerr Solution}
\label{review}

Before proceeding to the uniqueness proof, we briefly review the near horizon geometry of an extremal Kerr black hole \cite{Zaslavsky:1997uu,Bardeen:1999px}. Since the horizon of an extremal Kerr black hole is infinitely far away (in spacelike directions) from events outside the horizon, one can extract a limiting geometry by taking a certain scaling limit.  The general Kerr metric is labeled by two parameters, a mass $M$ and  angular momentum $J=Ma.$  
 In Boyer-Lindquist coordinates $(\rh,\th,\theta,\phih)$, the metric takes the form
\be\label{kerr}
ds^2=-e^{2\nu}d\th^2+e^{2\psi}(d\phih + \tilde \omega d\th)^2+\Sigma(d\rh^2/\Delta+d\theta^2)
\ee
where
\be
\Sigma=\rh^2+a^2\cos^2\theta, \quad \quad \Delta=\rh^2-2M\rh+a^2
\ee
\be
e^{2\nu}=\frac{\Delta\Sigma}{(\rh^2+a^2)^2-\Delta a^2\sin^2\theta},~
e^{2\psi}=\Delta\sin^2\theta e^{-2\nu},~
\tilde \omega = -\frac{2M\rh a}{\Delta\Sigma}e^{2\nu}.
\ee

Consider the extremal solution,  $a=M$. Defining a one-parameter family of new coordinate systems
\be
\rh=M+\lambda r,\qquad \th={t\over \lambda},\qquad \phih=\phi+{t\over 2M\lambda} \label{nhekscaling}
\ee
and taking the scaling limit $\lambda\rightarrow 0$ yields
\be
ds^2=\left(\frac{1+\cos^2\theta}{2} \right)\left[ -{r^2\over r_0^2} dt^2+{r_0^2\over r^2}dr^2+r_0^2d\theta^2\right]+\frac{2r_0^2\sin^2\theta}{1+\cos^2\theta}\left(d\phi+{ r\over r_0^2} dt\right)^2\, .\label{nhekmetric}
\ee
with $r_0^2=2M^2$.   The shift in $\phi$ is needed so that $\partial/\partial t$ is null on the horizon $r=0$. Since $\theta$ is not changed, the rotation axis ($\theta = 0,\pi$) in this limiting geometry agrees with the axis in Kerr near the horizon. This spacetime is known either as the extremal Kerr throat or as the Near-Horizon Extreme Kerr (NHEK) geometry.   It has recently attracted considerable attention in connection with a proposed Kerr/CFT correspondence \cite{Guica:2008mu}. For fixed $\theta$, the term in square brackets becomes the metric on $AdS_2$ in Poincar\'e coordinates.  
In fact, the NHEK geometry
 inherits all the isometries of AdS${}_2$.  It has an $SL(2,R) \times U(1)$ isometry group.

We are now ready to state and prove our uniqueness theorem for the extremal Kerr solution:

\bigskip
{\bf Theorem 1:} The only stationary,  rotating, asymptotically flat (analytic) vacuum solution with a single  degenerate horizon is  the extremal Kerr black hole.
\bigskip

{\it Proof:}  We follow the approach in \cite{Hollands:2007aj,Hollands:2008fm} which is based on earlier work by Mazur \cite{Mazur:1982db}. Many aspects of our proof are identical to the one proving uniqueness of nonextremal black holes. For those aspects,  we will just give the main ideas. For technical details, we refer the reader to \cite{Hollands:2007aj,Hollands:2008fm,Chrusciel:2008js}.  


It has recently  been shown that stationary (analytic) extremal  black holes must  be axisymmetric if they are rotating \cite{Hollands:2008wn,Chrusciel:2008js}.\footnote{The unlikely possibility of a stationary (but not static)  extremal black hole with zero angular velocity has not yet been ruled out.}   It therefore suffices to consider stationary, axisymmetric metrics.  
Such metrics  can always be written in  Weyl-Papapetrou form
\be\label{papapetrou}
 ds^2 = -{\rho^2 \over f} dt^2 + f(d\phi + \omega dt)^2 + e^{2\gamma}(d\rho^2 + dz^2)
 \ee
 where $f, \omega,  \gamma$ are functions of $\rho$ and $z$ only. Given a solution for $f$ and $\omega$, $\gamma$ is then determined in terms of them by first order equations. Rather than work with $\omega$, it is convenient to introduce the potential $\chi$ for the twist of the $\xi = \partial/\partial\phi$ Killing field:
 \be\label{twist}
 d\chi = *(\xi \wedge d\xi).
 \ee
 The twist potential (and Weyl-Papapetrou coordinates) are globally well defined in the domain of outer communication.
 
 A key role in the proof will be played by the following  2 x 2 matrix constructed from the norm and twist of $\xi$:
 \be
 \Phi = {1\over f}\left(\begin{array}{cc} 1 &  - \chi \\-\chi & f^2 + \chi^2\end{array}\right) .
 \ee
The matrix $\Phi$ is symmetric, has positive trace and unit determinant. It is therefore positive definite and can be written $\Phi = S^T S$ for some matrix $S$  with $\det S = 1$. The equation satisfied by $\Phi$ is most easily expressed by viewing $\rho$ and $z$ as cylindrical coordinates in an auxiliary flat Euclidean ${\bf R}^3$, with derivative $\nabla_i$. Viewing $\Phi$ as a rotationally invariant matrix in this space, the vacuum Einstein equation implies
 \be\label{Phieq}
\nabla^i (\Phi^{-1} \nabla_i \Phi) = 0
\ee
where this equation holds everywhere except possibly the axis $\rho =0$. 

Suppose we have two axisymmetric solutions $\Phi_1$ and $\Phi_2$ to this equation with the same angular momentum. Set
\be
\sigma = \Tr(\Phi_2 \Phi_1^{-1}) - 2 .
\ee
In terms of the norm and twist of the Killing field,
\be\label{posform}
\sigma = {(\chi_1- \chi_2)^2 + (f_1 - f_2)^2\over f_1 f_2} .
\ee
In this form, it is clear that $\sigma\ge 0.$
One can show that away from the axis $\rho =0$, $\sigma$ satisfies the following ``Mazur identity"
\be \label{mazur}
\nabla^2 \sigma = \Tr (N_i^T N^i),
\ee
where 
\be
N_i = S_2(\Phi_2^{-1} \nabla_i\Phi_2 - \Phi_1^{-1}\nabla_i\Phi_1) S_1^{-1}.
\ee
 Note that the right hand side of (\ref{mazur}) is nonnegative.
    
The requirements that $\nabla^2 \sigma \ge 0$ and $\sigma \ge 0$ impose strong constraints on $\sigma$. If we can show that $\sigma $ is globally bounded on ${\bf R}^3$ (including the axis) and vanishes at infinity then $\sigma $ must vanish everywhere \cite{Weinstein,Weinstein:1995tg}. This, in turn, implies that
$\Phi_1 = \Phi_2$ and hence the two solutions agree. 

We now show that $\sigma$ is indeed globally bounded. 
Since the key step is the behavior near the horizon, we consider this first. It was shown in \cite{Kunduri:2008tk} that the near horizon geometry of any extremal rotating vacuum black hole is given by the NHEK solution (\ref{nhekmetric}).  To put this into standard form (\ref{papapetrou}), note that the $(r,\theta)$ part of the metric is conformal to $dr^2 + r^2 d\theta^2$, so if one sets
\be\label{defrho}
\rho = r\sin\theta, \quad z=r\cos\theta,
\ee
 then (\ref{nhekmetric}) takes the form (\ref{papapetrou}). In other words, the radial coordinate in (\ref{nhekmetric}) is the standard radial coordinate in the auxiliary space ${\bf R}^3$. In particular,  the horizon corresponds to the origin of this space.  
 
Since the angular momentum can be expressed in terms of a Komar integral involving $\xi$, the value of $J$ in the NHEK metric must agree with the value computed at infinity. This fixes the free parameter $r_0$ in  (\ref{nhekmetric}) to be $r_0^2 = 2J$. For the NHEK geometry, the twist potential is
\be\label{chinhek}
\chi_{NHEK} = -{4r_0^2\cos\theta\over 1+\cos^2\theta}
\ee
and $\Phi$ is a function of $\theta$ only\footnote{The large symmetry group of the NHEK geometry ensures that all geometric quantities are functions of $\theta$ only.}:
\be\label{phinhek}
\Phi_{NHEK} = {1\over 2r_0^2\sin^2\theta} \left(\begin{array}{cc}{1+\cos^2\theta} & {4r_0^2\cos\theta} \\{4r_0^2\cos\theta}  & {4r_0^4(1+\cos^2\theta)}\end{array}\right).
\ee
So $\Phi_{NHEK}$ has a direction dependent limit at the horizon and diverges near the axis. 

To show that $\sigma$ remains bounded we consider the two terms in (\ref{posform}) separately. Since $f$ is the norm of the rotational Killing field, it must be smooth near the horizon.  The near horizon geometry must be given by the NHEK metric \cite{Kunduri:2008tk} with $r_0^2 = 2J$, so we have $f_1 = f_2( 1 + \alpha) $ where $\alpha$ is smooth and vanishes on the horizon.  It follows that the ratio $f_1/f_2$ goes to one everywhere on the horizon including the axis. This shows that in the auxiliary space ${\bf R}^3$, the second term in (\ref{posform}) vanishes at $r=0$, in a direction independent manner.
Now consider the first term. For $\theta \ne 0$ or $\pi$, $f$ is nonzero, and $\chi_1 - \chi_2 $ vanishes near the horizon since both $\chi_j$ must approach (\ref{chinhek}). So $\sigma$ again vanishes, at least for $\theta \ne 0,\pi$. Before discussing these points, we first consider the  axis away from the horizon.  

Smoothness near the axis requires $f_j = O(\rho^2)$, so the second term in (\ref{posform}) is bounded everywhere on the axis. Since the rotational Killing vector $\xi $ vanishes on the axis, its twist vector vanishes there and hence the twist potential $\chi$ is constant along the axis. The difference between $\chi$ on the $\theta = 0 $ axis and $\theta = \pi $ axis can be related to the Komar integral for $\xi$ and is given by $\Delta \chi = 8J$.  Hence given two extremal black holes with the same angular momentum, one can add a constant to the twist potential if necessary, so that $\chi_1 = \chi_2 $ on the axis.  Since $d\chi $ must vanish on the axis, $\chi_1 - \chi_2 =O(\rho^2)$ near the axis. It now follows from (\ref{posform}) that $\sigma$  remains bounded near the axis. (This argument is the same as in the nonextremal black hole case.) It remains to check that the coefficient of the $\rho^2$ term remains bounded as you approach the extremal horizon.  However it follows from (\ref{twist}) that the limit of $\partial^2_\rho \chi_j$ as $z \rightarrow 0 $ along the axis must approach $\partial^2_\rho \chi_{NHEK}$ which vanishes by (\ref{chinhek}).  So not only does the first term remain bounded as you approach the extremal horizon along the axis, it actually vanishes. We have thus shown that  $\sigma$ is bounded along the axis and vanishes at $r=0$  in a direction independent way. 

The treatment at infinity is the same as for nonextremal black holes, with the result that $\sigma$ vanishes asymptotically. For points off the axis, this follows from the fact that 
for any asymptotically flat spacetime, $f = \rho^2+$ subleading terms and $\chi$ remains bounded asymptotically. The axis requires a little more work but has the same conclusion. Hence $\sigma$ is globally bounded on ${ \bf R}^3$ and vanishes at infinity. Therefore it must vanish everywhere and $\Phi_1 = \Phi _2$.  This completes the proof.

\section{Uniqueness of the Extremal Kerr-Newman Solution}

We now show that the above result can be extended to extremal rotating and charged black holes. 
The near horizon geometry of the extremal Kerr-Newman solution is discussed in \cite{Zaslavsky:1997uu,Bardeen:1999px,Hartman:2008pb}.
It depends on a second parameter and smoothly interpolates between the solution 
(\ref{nhekmetric}) and $AdS_2\times S^2$. The Kerr-Newman
metric has exactly the same form as (\ref{kerr}), except that
$\Delta = {\td r}^2 -2 M \tilde r +a^2 +q^2$, where $q$ is the electric
charge  and the
$2 M \td r$ factor in $\tilde \omega$ is replaced by
$\td r^2 +a^2 -\Delta$.  The extremal limit
corresponds to $M^2 = a^2 +q^2$, and the horizon is
at $\td r = M$  with area $4\pi(M^2 +a^2)$. To obtain the throat metric, we can use the same scaling of $\td t,\td r$ as in (\ref{nhekscaling}), but the scaling of $\td\phi$ is modified to 
\be
\td \phi = \phi + {at\over r^2_0 \lambda}
\ee
where now $r_0^2 \equiv M^2 +a^2$.
 The near horizon geometry becomes
\be\label{nhekn}
 ds^2 = \left(1 -{a^2 \over r_0^2} \sin^2\theta\right) \left[ -{r^2\over r_0^2} dt^2 +
  {r_0^2 \over  r^2} dr^2 + r_0^2 d\theta^2\right]  
  \ee
$$+  r_0^2 \sin^2\theta \left(1 -{a^2 \over r_0^2} \sin^2\theta\right)^{-1} 
	\left( d\phi + {2arM\over r_0^4} dt\right)^2 .
	$$
Notice that when $a=0$, this metric reduces to $AdS_2\times S^2$ as expected. 
The Maxwell field in the extremal throat is $F=dA$ where the nonzero components of the vector potential are:
\be\label{Athroat}
A_\phi = {qaM\sin^2\theta\over M^2 + a^2\cos^2\theta}, \quad\quad A_t = {qr(M^2 - a^2\cos^2\theta)\over r_0^2(M^2 + a^2 \cos^2\theta)} .
\ee

We can now prove:

\bigskip
{\bf Theorem 2:} The only stationary,  rotating, asymptotically flat (analytic) Einstein-Maxwell solution with a single  degenerate horizon is  the extremal Kerr-Newman black hole.
\bigskip

The proof that stationary, rotating (analytic) extremal black holes must be axisymmetric in \cite{Hollands:2008wn} applies not just for vacuum spacetimes but also for Einstein-Maxwell solutions. So our solution must be axisymmetric and can be put into the Weyl-Papapetrou form (\ref{papapetrou}).
To prove Theorem 2, we will follow the original approach of Mazur \cite{Mazur:1982db,carter85}, which is based on the fact that the stationary, axisymmetric Einstein-Maxwell equations have an $SU(1,2)$ symmetry.\footnote{Setting the Maxwell field to zero, one recovers a uniqueness proof for Kerr based on the $SU(1,1)$ symmetry of the vacuum equations.  It will look slightly different from the proof given in the previous section, because we have used the equivalence of $SU(1,1) $ to $SL(2,R)$ to write that proof in terms of real matrices.} 
Given a stationary and axisymmetric Maxwell field, one can introduce two scalar potentials $E$ and $B$ as follows: Let $\xi$ be the rotational Killing field as before, and let $A$ be the vector potential, $F = dA$, in a gauge in which it is Lie derived by $\xi$. Similarly, we introduce a dual vector potential  ${}^* F = dC$ and pick a gauge in which $C$ is Lie derived by $\xi$. Then the scalar potentials are defined by $B= A_\mu \xi^\mu$ and  $E = C_\mu \xi^\mu$. We now define two complex Ernst potentials 
\be
\psi = E + i B, \quad \quad \epsilon = -f - |\psi|^2 + i\chi
\ee
where $f$ is the norm of the rotational Killing field $\xi$ as before. The definition of $\chi$ must be modified since the twist vector is no longer a gradient for Einstein-Maxwell solutions. Instead, we set 
\be
d\chi = *(\xi \wedge d\xi) + EdB - BdE .
\ee
The metric and Maxwell field are  completely determined in terms of $\epsilon,\psi$. 

Consider a complex three dimensional vector space with hermitian metric $\eta_{ab}$ with signature (1,2). So this is a complex analog of three dimensional Minkowski space. Let $v$ be the vector defined by 
\be
(v^0,v^1,v^2) = {1\over 2\sqrt f}(\ep -1,\ep+1,2\psi) .
\ee
Using a bar to denote the complex conjugate vector, one can easily check that $\eta_{ab} v^a \bar v^b = -1$, so $v^a$ is in fact a unit timelike vector. We now set
\be 
\Phi^{ab} = \eta^{ab} + 2 v^a \bar v^b .
\ee
This is a Hermitian $3\times 3$ matrix which is positive definite. In fact, one can view the second term as changing the sign of the time-time component of the original metric $\eta_{ab}$. $\Phi$ leaves the metric $\eta$ invariant in the sense that 
\be\label{phid}
\Phi^{am} \eta_{mn} \Phi^{nb} = \eta^{ab} .
\ee
Using $\eta_{ab}$ to raise and lower indices,
this implies that ${\Phi^a}_b {\Phi^b}_c = \delta^a_c$. It follows that
 $\Phi$ has unit determinant and so  defines an element of $SU(1,2)$. 

The equation satisfied by $\Phi$ is most easily expressed by again viewing $\rho$ and $z$ as cylindrical coordinates in an auxiliary flat Euclidean ${\bf R}^3$, with derivative $\nabla_i$. Viewing $\Phi$ as a rotationally invariant matrix in this space, the Einstein-Maxwell equations imply
 \be\label{Phieq}
\nabla^i (\Phi^{-1} \nabla_i \Phi) = 0
\ee
where, as before, this equation holds everywhere except possibly the axis $\rho =0$. Since $\Phi$ is positive, we can again write $\Phi = S^\dagger S$, and the proof now proceeds almost exactly as before.

Suppose we have two axisymmetric solutions $\Phi_1$ and $\Phi_2$ to (\ref{Phieq}). Set
\be
\sigma = \Tr(\Phi_2 \Phi_1^{-1}) - 3 .
\ee
In terms of our original quantities:
\bea\label{posformm}
\sigma =& {1 \over f_1 f_2} \big[(\Delta f)^2 + 2(f_1 + f_2)((\Delta E)^2 + (\Delta B)^2)+  [(\Delta E)^2 + (\Delta B)^2] ^2 \cr
&+(\Delta \chi + E_2 B_1 - E_1 B_2)^2\big]
\eea
where $\Delta f = f_1 - f_2 $ etc.
In this form, it is clear that $\sigma\ge 0.$
One can show that away from the axis $\rho =0$, $\sigma$ satisfies the following ``Mazur identity"
\be \label{mazurr}
\nabla^2 \sigma = \Tr N_i^\dagger N^i,
\ee
where 
\be
N_i = S_2 (\Phi_2^{-1} \nabla_i\Phi_2 - \Phi_1^{-1}\nabla_i\Phi_1) S_1^{-1}.
\ee
 Note that the right hand side of (\ref{mazurr}) is again nonnegative.
    
As in section 2, it suffices to show that $\sigma$ is globally bounded and vanishes at infinity \cite{Weinstein,Weinstein:1995tg}. 
Consider the  horizon first.  It was shown in \cite{Kunduri:2008tk} that the near horizon geometry of any extremal rotating and charged black hole is given by  (\ref{nhekn}).  The $(r, \theta)$ part of this metric is still conformal to $dr^2 + r^2 d\theta^2$
so one can use (\ref{defrho}) to put the metric into $\rho,z$ coordinates. 
The horizon again corresponds to the origin of ${\bf R}^3$.
 
For $\theta \ne 0, \pi$, $\Phi$ is finite in the limit $r\rightarrow 0$. If $\Phi_1$ and $\Phi_2$ are two solutions with the same charge and angular momentum,  the fact that they must agree near $r=0$ implies that $\Phi_1 = \Phi _2 +$ subleading terms.   Thus $\sigma \rightarrow 0$ in the limit $r \rightarrow 0$ for all $\theta \ne 0,\pi$.  Before discussing these points we consider the behavior of $\sigma$ on the axis away from the horizon.

The first term in  (\ref{posformm}) can be treated exactly as in the vacuum case with the result that it is bounded on the axis and vanishes as $r\rightarrow 0 $ for all $\theta$ including $\theta = 0,\pi$. We now consider the scalar potentials.   Since $B= A_\mu \xi^\mu$ and $A_\mu$ is globally well defined, $B=0$ on the axis. Smoothness requires $B = O(\rho^2)$, so the $(\Delta B)^2$ terms remain bounded on the axis. The dual vector potential $C_\mu$ is not globally defined since we have nonzero electric charge. Choosing a gauge so that the ``Dirac string" lies along the axis, we have that $E = C_\mu \xi^\mu$ is constant along the axis and $E (\theta = 0) - E(\theta = \pi) = 2q$. So given two solutions with the same charge, the values of $E_j$ along the axis will agree and $\Delta E = O(\rho^2)$. This ensures that the  $(\Delta E )^2$ terms will also be bounded. The mixed terms $E_jB_k$ in (\ref{posformm}) are also $O(\rho^2)$ and their contribution to $\sigma$  remains bounded. Finally, the potential $\chi$ behaves as in the vacuum case: It is constant along the axis, and the difference between its values on the two axes is $8J$. So two solutions with the same angular momentum will have $\chi_1 = \chi_2 + O(\rho^2)$ and the $(\Delta \chi)^2$ term is also bounded. 

Let us now consider the limit as we approach the extremal horizon along the axis. In the throat geometry:
\be
E_{throat} = {qr_0^2\cos\theta\over M^2 + a^2\cos^2\theta},\quad \quad
B_{throat} = {qaM\sin^2\theta\over M^2 + a^2\cos^2\theta}
\ee
As one approaches the extremal horizon along the axis, $\partial_\rho^2 B_j$ each   approach $\partial_\rho^2 B_{throat}$. Similarly $\partial_\rho^2 E_j$ each approach $\partial_\rho^2 E_{throat}$. So their difference vanishes. This shows that the limit of all terms in (\ref{posformm}) involving the electrostatic potentials vanish as one approaches the horizon along the axis. Similarly,  $\partial_\rho^2 \chi_j$ each approach the corresponding expression in the throat and hence the $(\Delta \chi)^2$ term vanishes. In short, all terms in the expression for $\sigma$ vanish as one approaches the horizon along the axis. 

One can again show that $\sigma$ vanishes at infinity along the same lines as in the nonextremal proofs. Hence $\sigma$ is globally bounded on ${ \bf R}^3$ and vanishes at infinity. Therefore it must vanish everywhere and $\Phi_1 = \Phi _2$.  This completes the proof.

\vskip .5 cm

{\bf Note added:} After completion of this work, we were informed of \cite{Meinel} which contains a proof of the uniqueness of extremal Kerr (but not Kerr-Newman) assuming axisymmetry. Their proof uses a different approach \cite{Neugebauer} from the one presented here.
\vskip 1cm
\centerline{\bf Acknowledgements}
\vskip .5 cm
It is a pleasure to thank S. Hollands, R. Wald, and G. Weinstein for discussions.
  This work was supported in part by the US National Science Foundation under Grant No.~PHY05-55669, and by funds from the University of California.


\begin{thebibliography}{99}

\bibitem{Hawking}
S. W. Hawking, ``Black holes in general relativity",
Commun. Math. Phys. {\bf 25} (1972) 152.

\bibitem{Carter}
B. Carter, ``Black hole equilibrium states",  in Black Holes (C. DeWitt and B. de Witt, eds.) Gordon and Breach, New York (1973).

\bibitem{Robinson:1975bv}
  D.~C.~Robinson,
  ``Uniqueness of the Kerr black hole,''
  Phys.\ Rev.\ Lett.\  {\bf 34} (1975) 905.
  
\bibitem{Mazur:1982db}
  P.~O.~Mazur,
  ``Proof Of Uniqueness Of The Kerr-Newman Black Hole Solution,''
  J.\ Phys.\ A  {\bf 15} (1982) 3173.
  
  \bibitem{Bunting}
  G. L. Bunting, ``Proof of the uniqueness conjecture for black holes", PhD Thesis, Univ. of New England, Armidale, N.S.W. (1983).
  
\bibitem{Hollands:2008wn}
  S.~Hollands and A.~Ishibashi,
  ``On the `Stationary Implies Axisymmetric' Theorem for Extremal Black Holes
  in Higher Dimensions,''
  arXiv:0809.2659 [gr-qc].
  
\bibitem{Chrusciel:2008js}
  P.~T.~Chrusciel and J.~Lopes Costa,
  ``On uniqueness of stationary vacuum black holes,''
  arXiv:0806.0016 [gr-qc].
  
\bibitem{Sudarsky:1992ty}
  D.~Sudarsky and R.~M.~Wald,
  ``Extrema of mass, stationarity, and staticity, and solutions to the Einstein
  Yang-Mills equations,''
  Phys.\ Rev.\  D {\bf 46} (1992) 1453.
  
\bibitem{Chrusciel:1993cv}
  P.~T.~Chrusciel and R.~M.~Wald,
  ``Maximal hypersurfaces in asymptotically stationary space-times,''
  Commun.\ Math.\ Phys.\  {\bf 163} (1994) 561
  [arXiv:gr-qc/9304009].
  

  
\bibitem{Friedrich:1998wq}
  H.~Friedrich, I.~Racz and R.~M.~Wald,
  ``On the Rigidity Theorem for Spacetimes with a Stationary Event Horizon or a
  Compact Cauchy Horizon,''
  Commun.\ Math.\ Phys.\  {\bf 204} (1999) 691
  [arXiv:gr-qc/9811021].
  
\bibitem{Kunduri:2007vf}
  H.~K.~Kunduri, J.~Lucietti and H.~S.~Reall,
``Near-horizon symmetries of extremal black holes,''
  Class.\ Quant.\ Grav.\  {\bf 24} (2007) 4169
  [arXiv:0705.4214 [hep-th]].
  

  
\bibitem{Kunduri:2008tk}
H.~K.~Kunduri and J.~Lucietti,
  ``A classification of near-horizon geometries of extremal vacuum black
  holes,''
  arXiv:0806.2051 [hep-th];
``Uniqueness of near-horizon geometries of rotating extremal AdS(4) black
  holes,''
  Class.\ Quant.\ Grav.\  {\bf 26} (2009) 055019
  [arXiv:0812.1576 [hep-th]].
  
  \bibitem{Hajicek}
  P. Hajicek,
  ``Three remarks on axisymmetric, stationary horizons,"
  Commun. Math. Phys. {\bf 36} (1974) 305.
  
\bibitem{Lewandowski:2002ua}
  J.~Lewandowski and T.~Pawlowski,
  ``Extremal Isolated Horizons: A Local Uniqueness Theorem,''
  Class.\ Quant.\ Grav.\  {\bf 20} (2003) 587
  [arXiv:gr-qc/0208032].
  
\bibitem{Zaslavsky:1997uu}
  O.~B.~Zaslavskii,
  ``Horizon/Matter Systems Near the Extreme State,''
  Class.\ Quant.\ Grav.\  {\bf 15} (1998) 3251
  [arXiv:gr-qc/9712007].
  
\bibitem{Bardeen:1999px}
 J.~M.~Bardeen and G.~T.~Horowitz,
 ``The extreme Kerr throat geometry: A vacuum analog of AdS(2) x S(2),''
 Phys.\ Rev.\  D {\bf 60} (1999) 104030.
 [arXiv:hep-th/9905099].
 
\bibitem{Guica:2008mu}
 M.~Guica, T.~Hartman, W.~Song and A.~Strominger,
 ``The Kerr/CFT Correspondence,''
 arXiv:0809.4266 [hep-th].




\bibitem{Hollands:2007aj}
  S.~Hollands and S.~Yazadjiev,
  ``Uniqueness theorem for 5-dimensional black holes with two axial Killing
 fields,''
  Commun.\ Math.\ Phys.\  {\bf 283} (2008) 749
  [arXiv:0707.2775 [gr-qc]].

\bibitem{Hollands:2008fm}
  S.~Hollands and S.~Yazadjiev,
 ``A uniqueness theorem for stationary Kaluza-Klein black holes,''
  arXiv:0812.3036 [gr-qc].

  
  
 
  
  
      \bibitem{Weinstein}
G. Weinstein, ``On the Dirichlet problem for harmonic maps with prescribed singularities", Duke Math. J. {\bf 77}  (1995) 135.
  
\bibitem{Weinstein:1995tg}
  G.~Weinstein,
  ``Harmonic Maps with Prescribed Singularities on Unbounded Domains,'' American Journal of Mathematics, {\bf 118} (1996) 689 
  [arXiv:dg-ga/9509003].



\bibitem{Hartman:2008pb}
  T.~Hartman, K.~Murata, T.~Nishioka and A.~Strominger,
  ``CFT Duals for Extreme Black Holes,''
  JHEP {\bf 0904} (2009) 019
  [arXiv:0811.4393 [hep-th]].
  
   \bibitem{carter85}
  B. Carter, ``Bunting Identity and Mazur Identity for Non-Linear Elliptic Systems Including the Black Hole Equilibrium Problem", Commun. Math. Phys. {\bf 99} (1985) 563.
  
  \bibitem{Meinel}
  R. Meinel, M. Ansorg, A. Kleinwachter, G. Neugebauer, and D. Petroff,
  {\sl Relativistic Figures of Equilibrium}, (Cambridge University Press, 2008) section 2.4.
  
  \bibitem{Neugebauer}
  G. Neugebauer,
  ``Rotating bodies as boundary value problems,"
  Ann.  Phys. (Leipzig) {\bf 9} (2000) 342; G. Neugebauer and R. Meinel,
  ``Progress in relativistic gravitational theory using the inverse scattering method,"
  J. Math. Phys. {\bf 44} (2003) 3407.
  

\end{thebibliography}
\end{document}